\documentclass[sigconf,10pt]{acmart}

\acmConference[ACM MobiCom '24]{The 30th Annual International Conference On Mobile Computing And Networking}{Fall, 2024}{Washington, D.C., USA}

\usepackage{graphicx} 
\usepackage{kotex}   
\usepackage{comment}
\usepackage{xspace}
\usepackage{bm}
\usepackage[normalem]{ulem}
\usepackage[linesnumbered,ruled,vlined]{algorithm2e}
\usepackage{multirow}
\usepackage{colortbl}
\usepackage{multicol}
\usepackage{graphics}

\usepackage{titlesec}
\titlespacing*{\subsubsection}{0pt}{0.1\baselineskip}{0.2\baselineskip}
\titlespacing*{\subsection}{0pt}{0.2\baselineskip}{0.2\baselineskip}
\titlespacing*{\section}{0pt}{0.4\baselineskip}{0.2\baselineskip}

\usepackage{color}
\definecolor{gray}{rgb}{0.4,0.4,0.4}
\definecolor{darkblue}{rgb}{0.0,0.0,0.6}
\definecolor{cyan}{rgb}{0.0,0.6,0.6}
\definecolor{maroon}{rgb}{0.5,0,0}
\definecolor
{darkgreen}{rgb}{0,0.5,0}
\definecolor{lightgray}{rgb}{0.8,0.8,0.8}
\definecolor{red}{rgb}{0,0,0}

\usepackage{xcolor}
\definecolor{codegreen}{rgb}{0,0.6,0}
\definecolor{codegray}{rgb}{0.5,0.5,0.5}
\definecolor{codepurple}{rgb}{0.58,0,0.82}
\definecolor{backcolour}{rgb}{0.95,0.95,0.92}

\usepackage{listings}
\lstdefinestyle{mystyle}{
    frame=tb,
    showstringspaces=false,
    columns=fullflexible,
    commentstyle=\color{codegreen},
    keywordstyle=\color{magenta},
    numberstyle=\tiny\color{codegray},
    numbers=none,
    stringstyle=\color{codepurple},
    basicstyle=\footnotesize,
    breakatwhitespace=false,         
    breaklines=true,                 
    captionpos=b,                    
    keepspaces=true,                            
    showspaces=false,                
    showstringspaces=false,
    showtabs=false,                  
    tabsize=2
}

\lstdefinelanguage{XML}{
    frame=tb,
    aboveskip=1mm,
    belowskip=1mm,
    showstringspaces=false,
    basicstyle=\footnotesize,
    morestring=[s]{"}{"},
    morecomment=[s]{?}{?},
    morecomment=[s]{!--}{--},
    commentstyle=\color{gray}\upshape,
    moredelim=[s][\color{black}]{>}{<},
    stringstyle=\color{black},
    morekeywords={index, id, text, description, UI_index},
    identifierstyle=\color{darkblue},
    keywordstyle=\color{cyan},
    tabsize=1,
    keepspaces=true,    
    showstringspaces=false,
    showtabs=false, 
    moredelim=**[is][\bfseries]{@}{@},
}
\definecolor{delim}{RGB}{20,105,176}
\definecolor{numb}{RGB}{106, 109, 32}
\definecolor{string}{rgb}{0.64,0.08,0.08}
\lstdefinelanguage{json}{
    frame=tb,
    aboveskip=1mm,
    belowskip=1mm,
    showstringspaces=false,
    basicstyle=\footnotesize,
    rulecolor=\color{string},
    breaklines=true,
    commentstyle=\color{gray}\upshape,    
    breakatwhitespace=true,
    morestring=[s]{"}{"},
    stringstyle=\color{black},
    keywordstyle=\color{string},
    morekeywords={name, desc, params, UI_index, UI_attrib, id, requires},
    moredelim=**[is][\bfseries]{@}{@},
    literate=
     *{0}{{{\color{numb}0}}}{1}
      {1}{{{\color{numb}1}}}{1}
      {2}{{{\color{numb}2}}}{1}
      {3}{{{\color{numb}3}}}{1}
      {4}{{{\color{numb}4}}}{1}
      {5}{{{\color{numb}5}}}{1}
      {6}{{{\color{numb}6}}}{1}
      {7}{{{\color{numb}7}}}{1}
      {8}{{{\color{numb}8}}}{1}
      {9}{{{\color{numb}9}}}{1}
      {\{}{{{\color{delim}{\{}}}}{1}
      {\}}{{{\color{delim}{\}}}}}{1}
      {[}{{{\color{delim}{[}}}}{1}
      {]}{{{\color{delim}{]}}}}{1},
}

\newcommand{\sys}{\textsf{\small MobileGPT}\xspace}

\newcommand{\sj}[1]{{\color[RGB]{0,50,200}\soln SJ: #1\solnend} }
\newcommand{\sk}[1]{{\color{olive}{\soln SK: #1\solnend}}}
\newcommand{\hy}[1]{{\color{pink}{\soln HY: #1\solnend}}}
\newcommand{\se}[1]{{\color{magenta}{\soln#1\solnend}}}
\newcommand{\secmt}[1]{{\color{magenta}{\soln SE: #1\solnend}}}
\newcommand{\iscmt}[1]{{\color{violet}{\soln IS: #1\solnend}}}
\newcommand{\red}[1]{{\color{red}{#1}}}

\title[MobileGPT]{MobileGPT: Augmenting LLM with Human-like App Memory for Mobile Task Automation}

\author{Sunjae Lee}
\orcid{0000-0002-2755-5511}
\affiliation{%
  \institution{School of Computing, KAIST}
  \country{Republic of Korea}
}
\email{sunjae1294@kaist.ac.kr}

\author{Junyoung Choi}
\orcid{0000-0003-3357-2710}
\affiliation{%
  \institution{School of Computing, KAIST}
  \country{Republic of Korea}
}
\email{joonchoi518@kaist.ac.kr}

\author{Jungjae Lee}
\affiliation{%
  \institution{School of Computing, KAIST}
  \country{Republic of Korea}
}
\email{dlwjdwo00701@kaist.ac.kr}

\author{Munim Hasan Wasi}
\affiliation{%
  \institution{School of Computing, KAIST}
  \country{Republic of Korea}
}
\email{munim@kaist.ac.kr}

\author{Hojun Choi}
\affiliation{%
  \institution{School of Computing, KAIST}
  \country{Republic of Korea}
}
\email{hchoi@kaist.ac.kr}

\author{Steven Y. Ko}
\affiliation{%
  \institution{Simon Fraser University}
  \country{Canada}
}
\email{steveyko@sfu.ca}

\author{Sangeun Oh}
\authornote{Co-corresponding authors}
\affiliation{%
  \institution{Korea University}
  \country{Republic of Korea}
}
\email{sangeunoh@korea.ac.kr}

\author{Insik Shin}
\authornotemark[1]
\affiliation{%
  \institution{School of Computing, KAIST}
  \institution{Fluiz Corp.,}
  \country{Republic of Korea}
}
\email{ishin@kaist.ac.kr}

\copyrightyear{2024}
\acmYear{2024}
\setcopyright{rightsretained}
\acmConference[ACM MobiCom '24]{The 30th Annual International Conference
on Mobile Computing and Networking}{November 18--22, 2024}{Washington
D.C., DC, USA}\acmBooktitle{The 30th Annual International Conference on Mobile Computing
and Networking (ACM MobiCom '24), November 18--22, 2024, Washington D.C.,
DC, USA}
\acmDOI{10.1145/3636534.3690682}
\acmISBN{979-8-4007-0489-5/24/11}

\begin{document}
\begin{abstract}

The advent of large language models (LLMs) has opened up new opportunities in the field of mobile task automation. Their superior language understanding and reasoning capabilities allow users to automate complex and repetitive tasks. However, due to the inherent unreliability and high operational cost of LLMs, their practical applicability is quite limited. To address these issues, this paper introduces \sys{}\footnote[1]{The system is available at: \url{https://mobile-gpt.github.io/}}, an innovative LLM-based mobile task automator equipped with a human-like app memory. \sys{} emulates the cognitive process of humans interacting with a mobile app---explore, select, derive, and recall. This approach allows for a more precise and efficient learning of a task's procedure by breaking it down into smaller, modular sub-tasks that can be re-used, re-arranged, and adapted for various objectives. 
We implement \sys{} using online LLMs services (GPT-3.5 and GPT-4) and evaluate its performance on a dataset of 185 tasks across 18 mobile apps. The results indicate that \sys{} can automate and learn \textit{new} tasks with 82.7\% accuracy, and is able to adapt them to different contexts with near perfect (98.75\%) accuracy while reducing both latency and cost by 62.5\% and 68.8\%, respectively, compared to the GPT-4 powered baseline.
\end{abstract}



\maketitle
\section{Introduction}
Today, we rely on a whole universe of mobile apps to handle tasks integral to our daily lives, ranging from communication to home security. Consequently, the demand for efficient task automation in the mobile environment has intensified, seeking to alleviate digital fatigue stemming from repetitive and complex digital interactions that often overwhelm many users today~\cite{digital_fatigue1, digital_fatigue2}.

Unsurprisingly, numerous methods have been explored to achieve task automation. API-based approaches like Siri~\cite{siri} and Google Assistant~\cite{google_assistant} allow users to interact with a mobile app's specific functionality through natural language conversation. However, these approaches demand significant coding efforts from developers, who must manually create new logic for each task.
Demonstration-based methods~\cite{sugilite, Amash, kite, Toby1} empower end-users to program custom automation scripts. Yet, due to their heavy reliance on human demonstrations, they face scalability challenges. Similarly, learning-based approaches~\cite{datadriven,mapping,mug,metagui,andenv} require extensive collections of human-annotated datasets, hampering the widespread application. 

Recently, LLM-based task automators~\cite{autodroid, autogpt, multion, hyperwrite, appagent}, equipped with high reasoning and generalization abilities~\cite{openai,llama,claude}, have been a game-changer. They enable task automation to be fully autonomous and generally applicable, sidestepping the labor-intensive manual development, demonstration, and training that previous methods required. However, this approach also comes with its own limitations. First, the inherent non-deterministic and unpredictable nature of LLMs can undermine the reliability and consistency of task automation. This is critical in mobile environments as many mobile tasks nowadays involve sensitive and private information.
Second, LLMs are costly, both in terms of budget and time. 
We observed that tasks that would take just over 30 seconds for a human could take more than two minutes for an LLM, and cost over a dollar each time they are performed.


To overcome the limitations of previous approaches, we introduce \sys{}, an innovative LLM-based mobile task automator augmented with an app memory capable of learning and recalling mobile tasks.
\sys{} is designed to accomplish following design goals: 
\textit{i) Accurate and Consistent}: it should perform tasks with high accuracy and consistency, ensuring that once a task is learned, it can be faithfully reproduced.
\textit{ii) Adaptability}: When revisiting a task, it should dynamically adjust its execution in response to varying contexts, rather than blindly replicating the previous executions.
\textit{iii) Efficiency}: \sys{} seeks to significantly reduce the cost and time involved in task automation, especially for tasks that are performed repeatedly.

In designing \sys{}, we drew inspiration from how humans decompose complex tasks into smaller sub-tasks to effectively learn and recall tasks~\cite{how_human_learn1, how_human_learn2, how_human_learn3, how_human_learn4}. Specifically, consider how humans learn new tasks using mobile apps: given an app screen, we first \textbf{1) explore} the candidate sub-tasks by analyzing screen interfaces and identifying their functionalities. Then, we \textbf{2) select} the most promising sub-task that can bring us closer to the goal. Lastly, we \textbf{3) derive} and execute the primitive actions required to complete the chosen sub-task---low-level actions such as click, input, and scroll. Once we have completed the task by repeating these 3 steps, it becomes part of our memory, allowing for easy \textbf{4) recall} and repetition of not only the task itself but also its involved \textit{sub-tasks.} For example, assume you have learned how to \textit{"Send a message to Bob."} Having learned this, not only can you easily adapt this knowledge to similar instructions like "Send a message to Alice," but also can apply it to execute new tasks such as "Read messages from John." This inherent human ability stems from our tendency to break down tasks into smaller \textit{sub-tasks} and encapsulate them in a modular, reusable format.

This human capacity to learn and recall memories at the unit of sub-tasks is what we aim to replicate with \sys{}. To achieve this, we address three key challenges: \textit{i) Accurate and reliable task execution}: To ensure accurate task execution and memory construction, we employ several prompting techniques to improve LLMs' accuracy and provide mechanisms for users to correct errors in case LLMs make mistakes. \textit{ii) Efficient memory storage:} To facilitate the use of memory, \sys{} efficiently parameterizes and stores task information in a hierarchical memory structure, in which tasks are decomposed into multiple function-call formatted sub-tasks and each sub-task is further decomposed into a sequence of primitive actions. \textit{iii) Flexible memory retrieval: } To ensure robust and cost-effective task automation in the dynamic landscape of mobile interfaces, \sys{} leverages pattern matching and in-context learning techniques to flexibly adapt actions involved in sub-tasks to varying contexts and interface changes.


We have implemented a prototype of \sys{} using online LLM services (GPT-3.5 and GPT-4). Our comparison study with state-of-the-art mobile task automators AutoDroid~\cite{autodroid} and AppAgent~\cite{appagent} demonstrates that \sys{} achieves a task completion rate of 82.7\% when executing a new task, outperforming these systems by 8\% and 15.3\% respectively. Furthermore, \sys{} achieves a near-perfect (98.75\%) success rate in adapting learned tasks to a new instruction with different task parameters. Our ablation study suggests that \sys{} achieves a 62.5\% reduction in task completion time and a 68.8\% decrease in LLM query costs when recalling learned tasks. 
A usability study with 23 participants demonstrates that \sys{}'s human-in-the-loop task repair mechanism enables users to interact intuitively with the task automator, allowing them to repair and collaboratively build upon the task automation process.


\section{System Overview}
\begin{figure*}[t]
    \centering
    \includegraphics[width=1.0\textwidth]{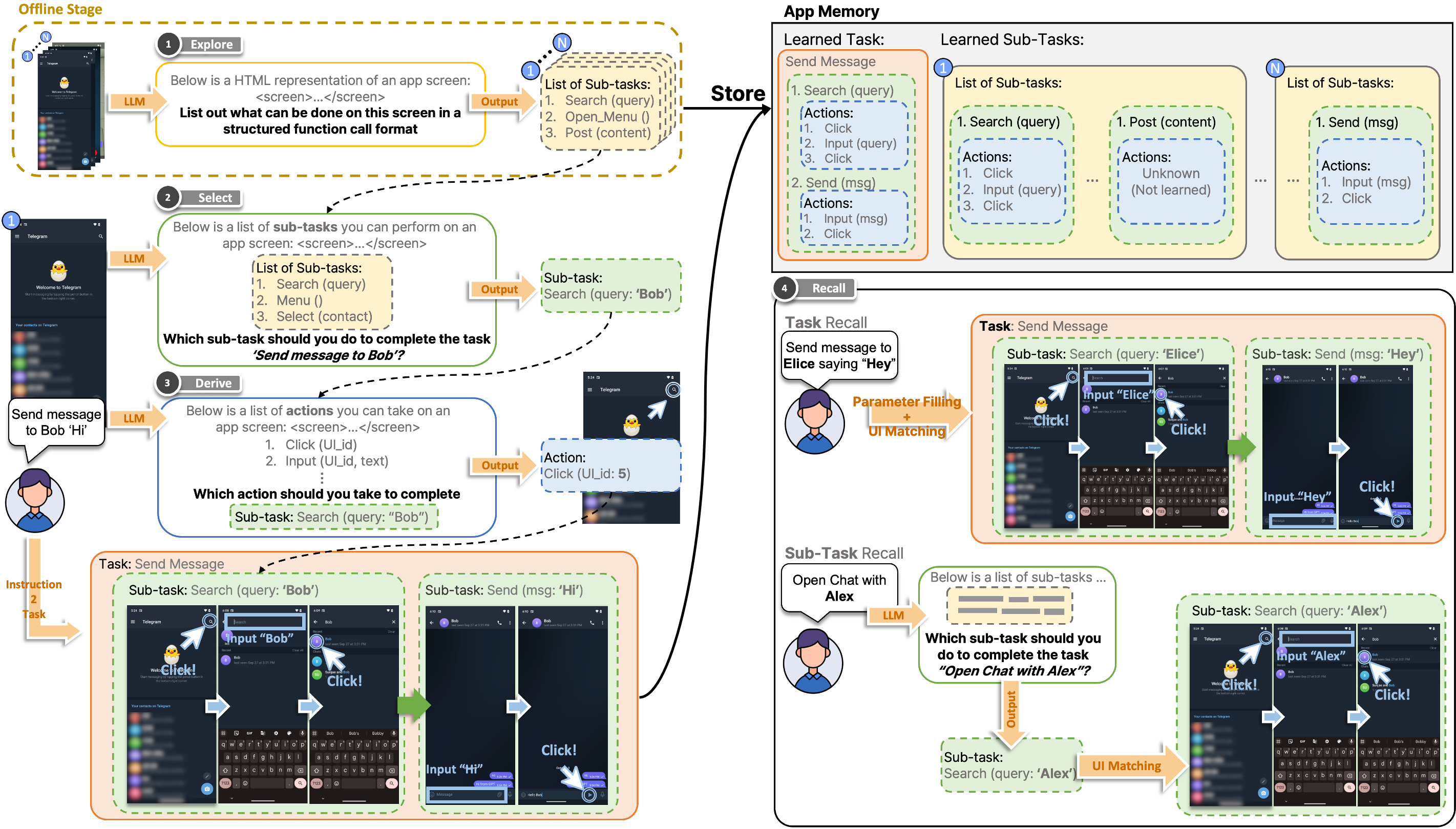}
    \caption{\sys{} workflow and system overview}
    \label{fig:workflow}
\end{figure*}



\subsection{System Workflow} 
Akin to the human cognitive learning process, \sys{} operates on a cycle of \textit{Explore-Select-Derive} phases to execute and learn new mobile tasks. This section gives a high-level overview of the system's workflow (\autoref{fig:workflow}).

\textbf{Explore.}
Prior to performing any tasks, \sys{} undertakes multiple \textit{explore} operations offline to pre-emptively collect and organize the functionalities of the app. To achieve this, \sys{} employs two software tools---random explorer~\cite{andenv} and user trace monitor~\cite{accessibility}---to visit and analyze as many app screens as it can during the offline stage. For each screen it visits, \sys{} asks the LLM to generate a list of sub-tasks available on the screen. This generated list is cached into memory to be used during the select phase upon receiving user instructions. App screens that have not been explored offline are analyzed on demand during the task execution. 



\textbf{Select.}
When a user issues a voice instruction, \sys{} retrieves the list of sub-tasks associated with the current screen from the memory. Then, it asks LLM which sub-task to perform in order to complete the user instruction. The selected sub-task is carried on to the Derive phase to be translated into a sequence of low-level actions.

\textbf{Derive.}
In the Derive phase, \sys{} prompts the LLM with a pre-defined set of low-level action (e.g., click, input, or scroll), and asks it to pick an action needed to accomplish the chosen sub-task. The action is then dispatched to the mobile device to be executed within the app. This phase is repeated until the app navigates to the next page, or the LLM explicitly indicates that the sub-task is completed. 

Upon finishing a sub-task, the system returns to the Select phase and selects the next sub-task to execute. This iterative process is continued until the Select phase indicates that the user's instruction has been fully executed. Then, the instruction is translated into a high-level task (e.g., \textit{"send a message to Bob" to "send message")} and saved in the memory in a hierarchical format, as illustrated in \autoref{fig:workflow}. 

\textbf{Recall.}
When the system is given an instruction while its memory is not empty, it checks if the instruction matches any previously learned task by comparing the instruction's high-level task representation with those stored in its memory. If a match is found, it executes the instruction directly from the memory, If no match is found, \sys{} cycles through Explore, Select, Derive to learn the new task. However, if the new task involves any known \textit{sub-tasks}, \sys{} can still reproduce them, akin to how humans transfer knowledge from one task to another. This facilitates faster execution and learning of new tasks.



\subsection{System Design}
To enable the workflow above, \sys{} addresses the following challenges:

\begin{itemize}
    \item [C1.] How to \textit{accurately and reliably execute} a task in the first try?
    \item [C2.] How to \textit{efficiently store} task executions?
    \item [C3.] How to \textit{flexibly recall} past task execution?
\end{itemize}
\textbf{C1. }
The first step towards learning a task is to correctly execute it the first time. \sys{} executes unknown tasks by leveraging multiple LLM queries to iterate through phases of Explore, Select, and Derive. However, given the complexity of mobile tasks and the unpredictability of LLM, we cannot guarantee the complete accuracy of these executions. For instance, when instructed to \textit{"find 5-star hotels,"} LLMs may simply type "5-star hotel" in the search field, whereas the expected behavior is to click the `5-star' filter option. 
Therefore, to address this inherent non-determinism and the unreliability of LLMs, \sys{} employs a dual-strategy correction mechanism that allows both the LLM and the user to fix errors in the execution process (more details in \autoref{sec:memory_construction}).


\textbf{C2.}
In many cases, tasks are not entirely independent of each other; they often share common sub-tasks. For instance, both `Send a message to Bob' and `Open chat with Alex'  involve the sub-task of locating a contact. If the execution procedures of the two tasks are stored separately at the task level, the sub-task they perform in common should be trained redundantly. To minimize such inefficiency, \sys{} employs a three-level hierarchical memory structure: \textit{tasks, sub-tasks, and actions}. This hierarchy enables \sys{} to access memory at the sub-task level, facilitating the sharing of past execution experience across different tasks.

\red{
Moreover, each task often comes with different variations. The task of sending a message can be specified as "send message to Bob" or "send Elice a message `Hey'." To address this range of variations, we need to properly parameterize the learned task. \sys{} achieves this by representing each subtask in a function-call format and the task as a series of these function calls (see \autoref{fig:workflow}). This approach allows for more granular parameterization, as \sys{} identifies parameters incrementally while proceeding with the task, including implicit parameters that are hidden in the instruction, such as the message content parameter in the instruction "send message to Bob."

}

\textbf{C3.} 
Even when repeating the same task or sub-task, the specifics of each step may vary depending on the context of execution. For example, searching for a contact in the context of "Send a message to Bob" involves entering and clicking "Bob" in the search page, whereas "Send a message to John" requires searching for "John" instead. Therefore, instead of blindly replicating past actions, \sys{} flexibly adapts its past executions to accommodate not only the intricate parameters of the task but also changes in screen content.
In addition, in case direct adaptation falls short, \sys{} leverages the few-shot learning capability of LLMs to guide them in generating responses that are both consistent and deterministic throughout multiple trials of the task (more details in \autoref{sec:recall}).
\section{Accurate and Reliable Task Execution}
\label{sec:memory_construction}
LLMs have been reported to solve complex tasks more accurately by breaking them down into smaller sub-tasks~\cite{cot, plan}. \sys{} adopts this strategy by hierarchically decomposing tasks into sub-tasks according to the Explore-Select-Derive phases. This section outlines how \sys{} effectively navigates each of these phases, while also addressing the challenges posed by the inherent unreliability of LLMs. For clarity, this section describes each phase under the assumption that \sys{}'s memory is empty. 





\subsection{Prompting Mobile Screen to LLM}
\label{sec:screen_representation}

The initial step in applying LLMs for mobile tasks involves converting mobile screens into text representation. Drawing upon previous research~\cite{enabling} showing that LLMs comprehend Graphical User Interfaces (GUIs) better when presented in HTML syntax, we convert mobile screens into a simplified HTML representation. 

We begin by extracting the screen's layout information using Android Accessibility Service~\cite{accessibility}. This information includes a hierarchical relationship between UI elements and various attributes (e.g., class\_name, text, descriptions) and properties (e.g., clickable, editable) that describe the functionality and appearance of the UI. To make the layout file succinct, we prune UI elements that are neither interactive nor carry significant semantic attributes (e.g., empty layout container). Then, we map each remaining UI element into an HTML element, where the UI's text and description attributes serve as the content of the HTML element, and the UI's interactive property is translated into the corresponding HTML tag. For instance, the `<button>' tag is used for clickable UIs, `<input>' for editable UIs, `<scroll>' for scrollable UIs, and so on. Lastly, we assign each HTML element with a unique index number, which serves as a communication link between \sys{} and LLMs to specifically refer to and interact with each element on the screen. This process not only ensures a cleaner, more relevant representation of app screens but also significantly reduces the number of tokens by an average of 84.6\%. Throughout this paper, we will refer to this HTML representation of an app screen as the \textit{"screen representation."}

\subsection{Explore, Select, and Derive} 
\label{sec:explore_select_and_derive}
We generate the screen representation every time there is a change in the screen. This screen representation is then used throughout the process of the Explore, Select, and Derive phases.

\textbf{Explore. }
The goal of the \textit{explore} phase is to generate a list of actionable sub-tasks for a given screen. Each sub-task represents an individual operation or functionality that the screen provides. 
To accomplish this, \sys{} prompts the LLM with the screen representation and asks it to enumerate sub-tasks in a structured function call format with the following information included: 1) sub-task name and description, 2) parameter names and descriptions, and 3) index of relevant UI elements. For example, possible sub-tasks for Telegram's initial app screen (\autoref{fig:workflow}) include:
\lstset{style=mystyle}
\lstset{language=json}
\begin{lstlisting}
1. { name: "Search", desc: "Search for a contact", params: {"query": "who are you looking for?" }, UI_index: 3 }
2. { name: "Open_Menu", desc: "Open menu", params: {}, UI_index: 7 }
\end{lstlisting}

Note that, unlike other phases, the explore phase operates independently of user instructions. Hence, we conduct this phase \emph{offline}, before performing the user's designated task. We use tools like random explorer~\cite{andenv}, which performs random UI actions to navigate through app screens, and user trace monitor~\cite{accessibility}, which tracks users' app usage in the background. These tools allow for the proactive collection of sub-tasks related to each app screen. While the current implementation of \sys{} conducts this offline phase locally on the user's device, it could be offloaded to a server to enhance efficiency.


\textbf{Select. }
During the \textit{select} phase, \sys{} prompts the LLM with \textit{i)} the user instruction, \textit{ii)} the current screen representation, and \textit{iii)} a list of available sub-tasks. \red{The list includes both the sub-tasks identified during the \textit{explore} phase and a predefined set of global sub-tasks common to all screens such as `Read\_Screen' for answering the user's question based on the current screen content and `Finish' for indicating task completion.} The LLM then picks the sub-task most pertinent to the instruction and fills in its required parameters. For instance, given the instruction "Send a message to Bob", the output of the first select phase would be: 
\begin{lstlisting}[mathescape=true]
{ name: "Search", description: "Search for a contact", parameters: {"query": $\bm{"Bob"}$ }, UI_index: 3 }
\end{lstlisting}
When parameter values are unknown, \sys{} asks the user for the missing information.
This allows for interactive communication between the user and the system, enabling task automation even when the user's instruction is not entirely clear.

After a sub-task has been selected and its parameters filled, \sys{} checks if the sub-task has already been learned (i.e., present in the memory) in the context of another task. If the sub-task is known, \sys{} executes the sub-task directly from the memory, bypassing the Derive phase. Otherwise, it proceeds to the Derive phase.

\textbf{Derive. }
During the \textit{derive} phase, \sys{} incrementally derives and executes low-level actions to accomplish the sub-task selected during the \textit{select} phase.
Specifically, \sys{} prompts the LLM with \textit{i)} the sub-task to execute, \textit{ii)} the current screen representation, and \textit{iii)} a pre-defined list of low-level actions---click, input, scroll, long-click---. Then, the LLM selects one of the actions from the list along with the index of the target UI element on which the action needs to be performed. For instance, if the index of the search button is 5, the output of the first \textit{derive} action for the sub-task \textit{`Search'} would be \textit{\textbf{click}(ui\_index=5)}. The action is then dispatched to the mobile device to be executed within the app. \sys{} repeats this process until there is a transition in the app screen or the LLM explicitly indicates that the sub-task is completed. 
\red{
Afterward, \sys{} returns to the Select phase to choose the next sub-task, and alternates between the Select and Derive phases until the LLM \textit{selects} the subtask \textit{`Finish'}, indicating that the user's instruction has been completed.
}

\subsection{Dual Strategy Failure Handling}
\label{sec:construc_failure_handling}



Despite their high reasoning abilities, LLMs sometimes show inconsistent and erroneous behavior. To address this, \sys{} employs a dual-strategy correction mechanism. 

\textbf{Self-Correcting through feedback generation. }
Previous studies~\cite{critic,self_feedback} indicate that language models can self-correct when given appropriate feedback. 
To facilitate this in \sys{}, we have heuristically identified two main types of commonly occurring errors and developed a rule-based self-feedback generation module. Upon detecting an error, \sys{} generates appropriate feedback and appends it at the end of the next prompt. This guides LLMs in refining their approach and self-correcting the ongoing execution.

The first type of error occurs when the LLM incorrectly attempts to interact with the UIs. Feedback for this error includes:\textit{"There is no UI with index $i$"} and \textit{"The UI is not (clickable)."} These errors can be easily detected as they result in a failure when executing the action. The second type of error is when LLM gets stuck in a loop. For example, endlessly scrolling through a YouTube video list or trying to scroll when already at the bottom of a list. Feedback for this error includes: \textit{"There is no change in the screen."} and \textit{"You have looped the same screens X times."} These errors can be detected by monitoring the screen changes and tracking the visited app screens.

\textbf{Human-in-the-loop (HITL) task repair. }
\begin{figure}[t]
    \centering
    \includegraphics[width=0.47\textwidth]{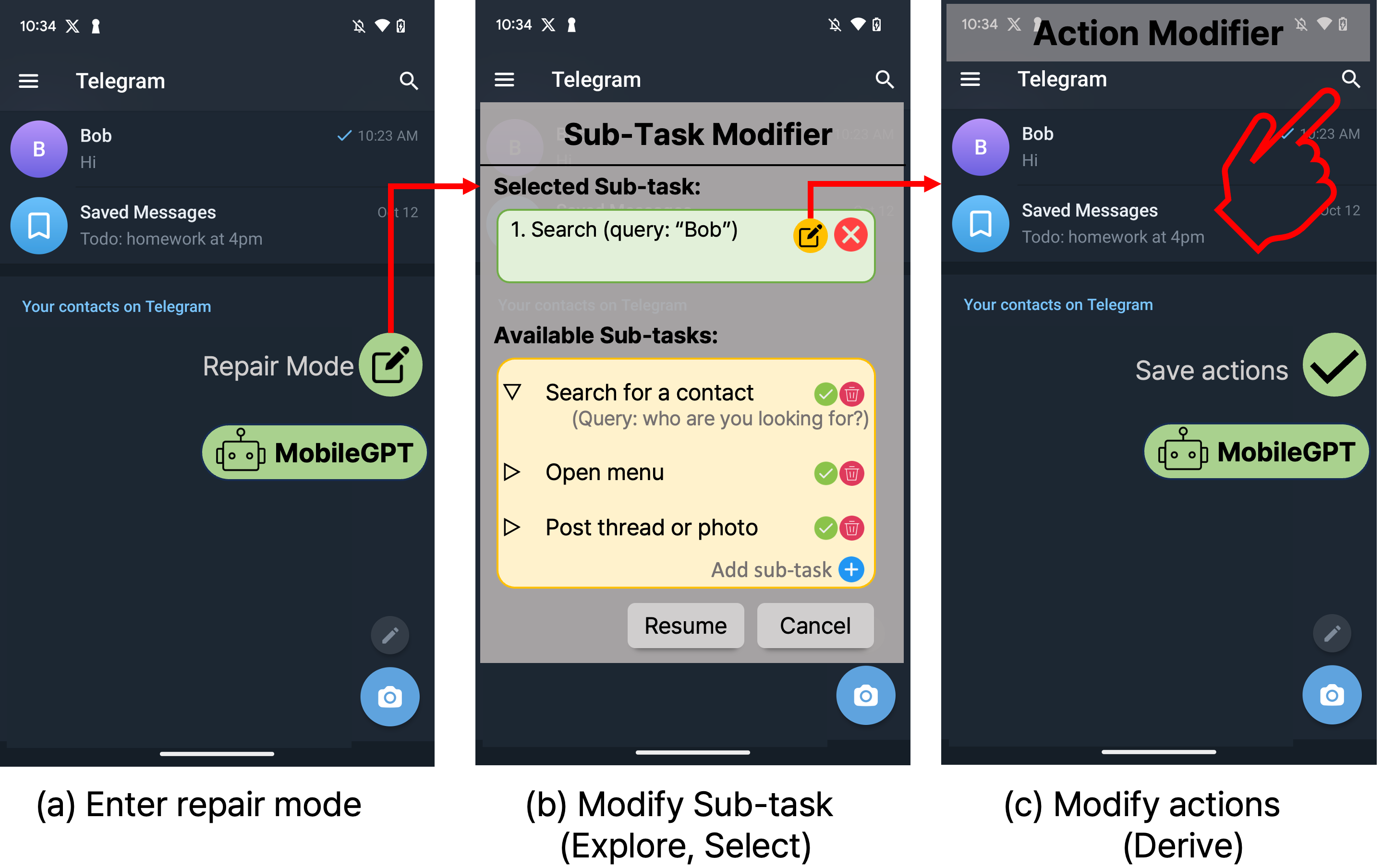}
    \caption{Illustration of HITL task repair mechanism}
    \label{fig:repair}
\end{figure}
In case self-correction falls short, \sys{} provides mechanisms for users to correct the LLM's mistakes themselves. In \sys{}, there are three major points of potential LLM failure: Explore, Select, and Derive. \textit{i)} \textit{Explore} fails when sub-tasks are missing from the list of available sub-tasks, \textit{ii)} \textit{Select} fails when the LLM chooses incorrect sub-tasks or inputs wrong parameters, and \textit{iii)} \textit{Derive} fails when incorrect actions are derived and executed. \sys{} provides specific repair mechanisms for each failure type\footnote[1]{see \url{https://mobile-gpt.github.io} for demo video.}.

\autoref{fig:repair} is a simplified illustration of the repair mechanism. As illustrated, users can enter the repair mode anytime during the task execution by clicking on the \sys{}'s floating button. Upon entering the repair mode, \sys{} pauses its execution and hands control over to the user. The user can then perform repairs directly on the current screen or navigate back to previous screens to re-do mistakes made by \sys{}. Note that certain critical actions, such as sending a message or deleting a contact, cannot be undone. In such cases, users would have to manually restore the app to its original state. Upon identifying the screen needing repair, users can \textit{i)} add or remove sub-tasks from the list of available sub-tasks (Explore), \textit{ii)} change the selected sub-task and its parameters (Select), or \textit{iii)} edit the actions involved in the sub-task by demonstrating a sequence of actions, similar to the programming by demonstration techniques (Derive). Additionally, \sys{} assists users in locating the point of repair by providing a detailed visual summary of the task execution both in terms of sub-tasks and actions. This allows users to pinpoint the failure point even if they were not closely monitoring the execution. 

A key advantage of \sys{}'s HITL repair is that users can return control back to the \sys{} after fixing an error. This seamless transition requires a mutual understanding between the users and the LLMs, so that LLMs can recognize the corrections made and refine their approach accordingly. \sys{}'s design---organizing tasks as a sequence of sub-tasks rather than low-level actions---plays a pivotal role. The natural language descriptions of sub-tasks not only help in delivering the LLM's current progress to the user but also effectively capture the users' intentions behind their repair. For example, when the user corrects actions involved in the sub-task `Search', the correction is encapsulated in a feedback \textit{"User repaired how to: Search for a contact"} and appended to the next prompt. This feedback enables the LLM to grasp the intent behind the user's repair and proceed accordingly. Without such contextual information, LLMs could repeat the same error or actions already performed by the user.

\section{Hierarchical App Memory}
\label{sec:memory_architecture}
\sys{}'s memory architecture resembles how humans get familiar with the app---accumulate knowledge about the app as a whole, not just the tasks themselves. This section outlines how \sys{} systematically archives the results of the Explore, Select, and Derive phases, and utilizes them for future task executions.

\begin{figure}[t]
    \centering
    \includegraphics[width=0.47\textwidth]{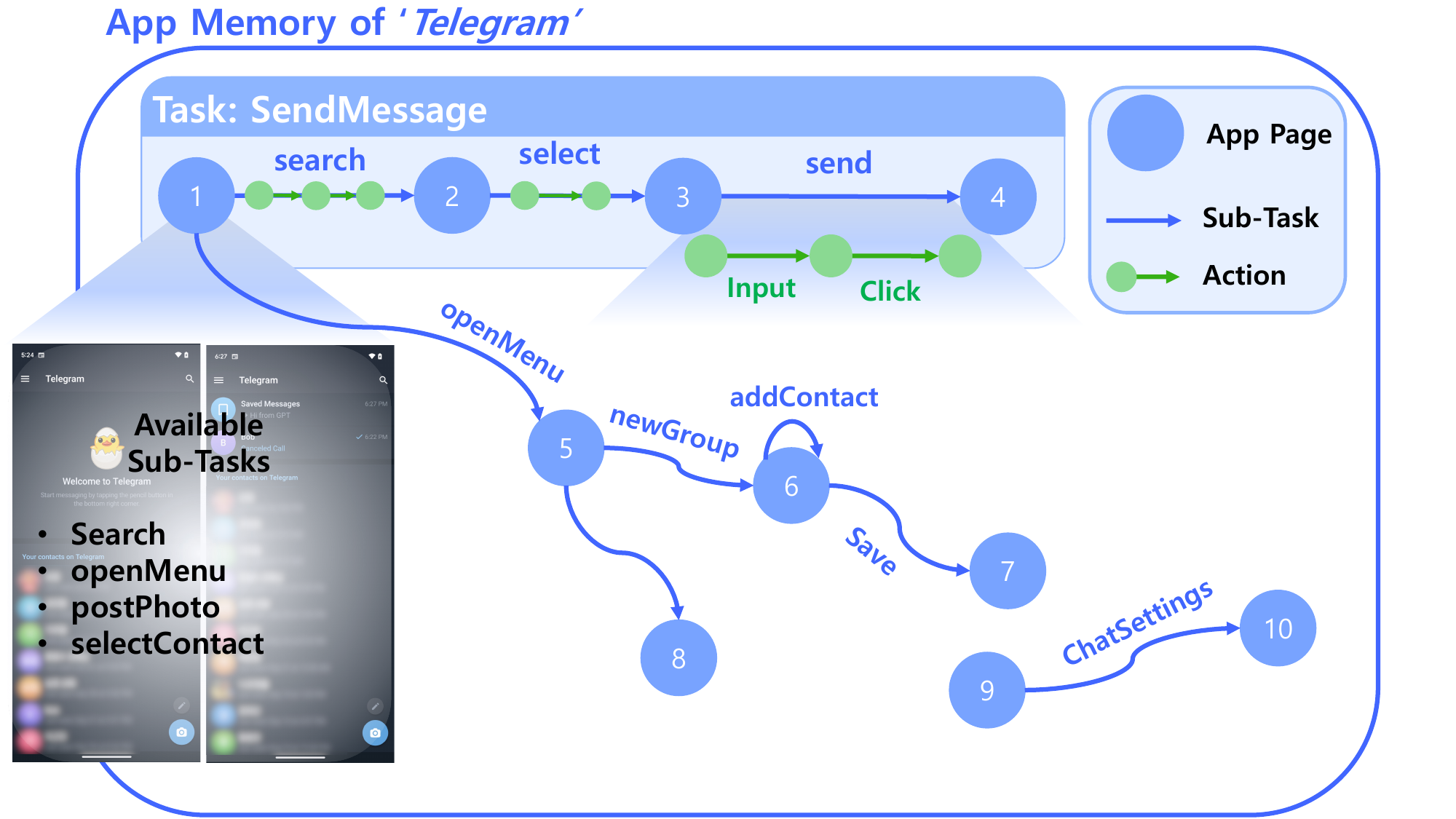}
    \caption{Transition graph of the app `Telegram'}
    \label{fig:memory_architecture}
\end{figure}

\subsection{Memory Structure}

\label{sec:memory_design}

\sys{} organizes its memory in the form of a transition graph that encapsulates the following key information: i) available sub-tasks of each app screen \textit{(Explore)}, ii) sequence of sub-tasks involved in each task \textit{(Select)}, and iii) how to perform each sub-task \textit{(Derive)}. \autoref{fig:memory_architecture} illustrates an example of this graph for the app `Telegram.' Note that these graphs exist per app, not per task, meaning that tasks within the same app all share the same graph. 

\begin{figure}[t]
    \centering
    \includegraphics[width=0.47\textwidth]{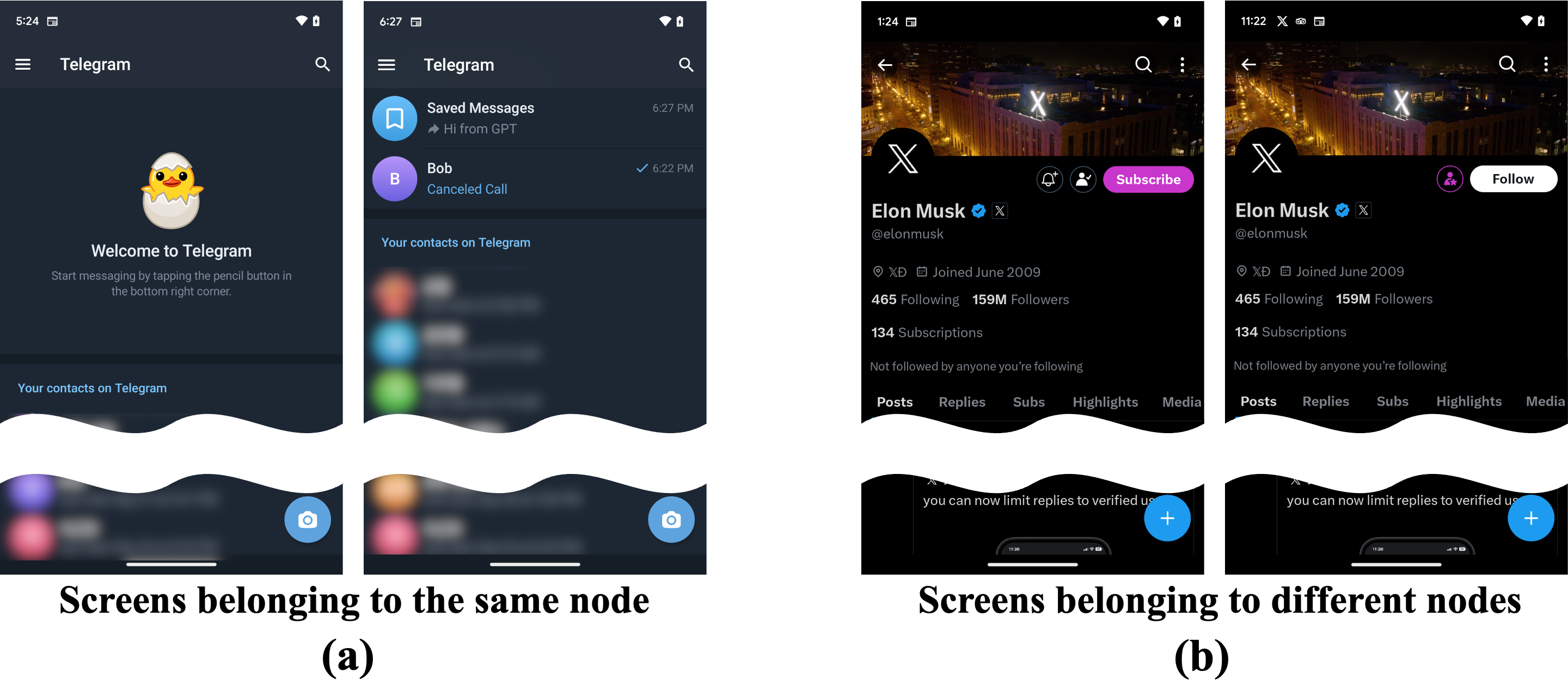}
    \caption{Examples of app pages: Screens in (a) belong to the same app page because they provide the same functionalities. Screens in (b) belong to different app pages because they provide different functionalities (the left is for `Subscribe' but the right is for `Follow').}
    \label{fig:screen_examples}
\end{figure}

\textbf{Node.}
Each node in the transition graph symbolizes an app \textit{page}---a particular state within an app that offers a unique set of functionalities (i.e., sub-tasks). \autoref{fig:screen_examples} illustrates examples of app pages; Note that screens with different visual appearances can belong to the same page (\autoref{fig:screen_examples} a), while screens that look similar (\autoref{fig:screen_examples} b) could belong to different pages, depending on their functionalities. 

To this end, \sys{} uses a list of sub-tasks to represent each page node (e.g., node 1 in \autoref{fig:memory_architecture}), categorizing screens that share the same list of sub-tasks as the same node. This facilitates the sharing of sub-task knowledge across a wider variation of app screens, ensuring that a sub-task learned on one screen can be applied to another, even if the two screens do not look exactly alike. For example, once \sys{} learns how to `Follow' in Elon Musk's profile page, it can use the same knowledge to `Follow' Mark Zuckerberg.

\textbf{Edge.}
Transitions between nodes are defined by sub-tasks, where each sub-task consists of a sequence of primitive actions that outline the steps for executing the sub-task. This hierarchical structure enables \sys{} to efficiently reproduce learned sub-tasks by following the sequence of actions involved. \sys{} establishes a new edge when it undergoes the Derive phase to execute previously unknown sub-tasks. Impotantly, these sub-task edges are not confined to individual tasks but are shared across tasks.


\textbf{Task. }
After \sys{} successfully completes the user instruction, it stores the task as a collection of node and sub-task pairs. This representation effectively indicates which sub-task needs to be executed on which page. Consequently, \sys{} can bypass all three phases of learning---Explore, Select, and Derive when executing previously learned tasks. Additionally, it facilitates state-agnostic task execution; Regardless of which app page the app is currently on, \sys{} can find which sub-task to execute.

\subsection{Sub-task based Screen classification}
\label{sec:memory_utilization}
Since an app page (i.e., a node) is a group of app screens that share the same list of sub-tasks, our memory architecture requires an accurate classification of app screens based on their functionalities. However, traditional screen classification methods~\cite{webui, screen2vec} are not suitable for this, as they focus on the screen's appearance, rather than its functionalities. 

A brute-force approach to classify screens by their functionalities would be to run the Explore phase (i.e., extract a list of available sub-tasks) each time we encounter a new screen and fuzzy match with those in the memory. However, this would incur significant overhead and cost. Therefore, \sys{} introduces a lighter approach to a sub-task-based screen classification: instead of checking if a screen \textit{has} the same list of sub-tasks, it verifies if the screen can \textit{perform} those sub-tasks.

Specifically, during the Explore phase, \sys{} asks the LLM to return each subtask with relevant UI indexes (see \autoref{sec:explore_select_and_derive}). These UI elements serve as requirements for performing a specific sub-task. For example, if the LLM returns the following `Search' subtask during the Explore phase,
\lstset{style=mystyle}
\lstset{language=json}
\begin{lstlisting}
{ name: "Search", desc: "Search for a contact", params: {...}, @UI_index: 3@ }
\end{lstlisting}
the Search button, which corresponds to the UI index of 3 will be the key element for the subtask `Search'.
\lstset{language=XML}
\begin{lstlisting}
<button @UI_index=3@ id="search_button" description="Search"/> 
\end{lstlisting}


Then, each time \sys{} arrives at a new screen, it verifies if the current screen includes key UI elements required by each sub-task. In specific, we check if the screen representation has a UI that matches the attributes (e.g., ui\_id, description, and class\_name) of the key UI elements. If the screen representation has matching UIs for all of the node's sub-tasks, we categorize the screen under that node. 

If no matching node is found, \sys{} undergoes the Explore phase. However, before creating a new node, \sys{} calculates the cosine similarity between the screen's newly generated sub-tasks and those of existing nodes to ensure that no existing node has the same list of sub-tasks. If a match is found, we map the screen to the node and update its sub-task information without creating a new node. This double-check process minimizes the creation of redundant nodes, preventing already established sub-task edges from becoming obsolete.

In our evaluation, where \sys{} encountered and classified 269 app screens, this method showed only 3 false positives (misclassifying different app pages as the same node) and no false negatives (failing to find the correct page node). The result is significantly better compared to existing SotA text-based~\cite{screen2vec} and vision-based~\cite{webui} screen classification methods, which had 38, and 57 false positives, and 14, and 28 false negatives, respectively.

\section{Flexible Task Recall}
\label{sec:recall}
When recalling tasks or sub-tasks, the specific details of each action should be adjusted to the current context of its execution. To address this, \sys{} leverages the attribute-based pattern matching and LLM's few-shot learning capability~\cite{few_shot} to adjust and reproduce actions flexibly and accurately.

\subsection{Attribute-based Action Adaptation.}
\label{sec:adapting}
There are two cases where we need to adjust the action. 
The first case is when a task parameter changes. For instance, if the user instruction changes from \textit{"Send a message to Bob"} to \textit{"Send a message to Alice"}, we need to search and click for \textit{"Alice"} instead of \textit{"Bob"}. The second case is when there is an alteration in the screen's content. For example, in a contact page, the hierarchical position of a specific contact (i.e., index of the UI) may change if contacts are added or removed. To effectively handle both scenarios, \sys{} generalizes and adapts actions to both the parameters and the screen contents.

\textbf{Generalizing Actions. }
When learning a new sub-task, \sys{} stores actions in a generalized format to ensure that they are reusable across different contexts. This involves two steps: Screen Generalization and Parameter Generalization. For a given action (e.g., \textit{\textbf{click} (ui\_index=5)}), \sys{} first generalizes it against the current screen representation by locating the target UI and recording its key attributes (e.g., id, text, description). Then, each key attributes are generalized further against the task parameters by comparing their values. If the value of the attribute matches that of the subtask parameter, the attribute's value is replaced with the corresponding parameter name.
For example, given a subtask \textit{`\textbf{select}(contact\_name: "Bob")'} and a screen representation:
\lstset{language=XML}
\begin{lstlisting}
    <!-- Contact list-->
    <button index=5 id="contact" text="Bob"/>
    <button index=6 id="contact" text="Alice"/>
\end{lstlisting}
the action `\textit{\textbf{click} (ui\_index=5)}' gets generalized through following two steps:
\lstset{style=mystyle}
\lstset{language=Python}
\begin{lstlisting}[mathescape=true]
click(ui_index=5) $\rightarrow$ 1. click(id:"contact", text: "Bob")
                             $\rightarrow$ 2. click(id:"contact", text:"[contact_name]")
\end{lstlisting}



\textbf{Adapting Actions. }
When recalling the whole task, \sys{} simply replays the given sequence of sub-tasks, bypassing all Explore, Select, and Derive phases. Yet, we still need to fill in the parameters for each sub-task and adjust the involved action accordingly. To do so, before executing each sub-task, \sys{} queries LLM to slot-fill parameters based on the user instruction and the current screen representation. Similar to the Select phase, if parameter values are unknown (e.g., missing from the instruction), \sys{} asks the user for the information.

To adapt an action to a new context, we simply reverse the aforementioned two generalization steps based on the given sub-task parameters and the screen representation. Specifically, \sys{} substitutes parameter names with the actual parameter values and identifies the UI element that matches these updated attributes. For example, given the sub-task \textit{`\textbf{select}(contact\_name: "Alice")'}, the generalized action for clicking a contact gets adjusted as follows:
\lstset{style=mystyle}
\lstset{language=Python}
\begin{lstlisting}[mathescape=true]
click(id:"contact", text:"[contact_name]")
$\rightarrow$ 1. click(id:"contact", text: "Alice") $\rightarrow$ 2. click(ui_index=6)
\end{lstlisting}
This two-step generalization and adaptation technique allows us to correctly locate the target UI in response to the change in the task parameters and screen.

\textbf{Benefits.}
One advantage of identifying parameters at the sub-task level and using them for the action adaptation is its ability to handle ambiguities and incompleteness often found in real-world user instructions. A prior work~\cite{sugilite}, which generalizes actions based on the words in the instruction, fails to handle incomplete instructions like \textit{"Send a message to Bob,"} where the message content is missing, and implicit instructions like \textit{"Call the first contact from the recent call,"} where the contact name isn't explicitly stated. 

On the other hand, \sys{} effectively addresses this challenge by identifying and deriving information required to perform the task incrementally at the sub-task level. For example, when processing \textit{"Send a message to Bob,"}, \sys{} proactively asks for the missing message content before performing the subtask \textit{\textbf{send}(message\_content)}. Similarly, for \textit{"Call the first contact from the recent call,"}, \sys{} autonomously identifies the contact information by looking at the recent call list before executing the sub-task \textit{\textbf{call}(contact\_name)}. This allows \sys{} to generalize actions against a more granular and precise set of parameters even if the user instruction is not entirely clear.

\subsection{In-context Action Adaptation}
The attribute-based adaptation method is not a one-size-fits-all solution, as several factors can lead to failure. First, the target UI may not include any key attributes for generalization. Second, multiple UIs could share the same attributes. Third, UIs may change unexpectedly with app updates. 

Therefore, when the attribute-based adaptation fails, we resort to LLM to re-derive the action once again. However, LLMs are inherently non-deterministic. This means there is no guarantee they will reproduce a previously correct action. Worse yet, if the LLM had made a mistake in the past, it has a high probability of repeating the same error.

To address this issue, we leverage the few-shot learning capability of the LLM ~\cite{cot, few_shot} to produce more accurate and reliable responses. When querying the LLM, we present it with an example of how the action (i.e., one that rule-based adaptation failed) has been derived in the past. Specifically, the example includes a prior user instruction, an abbreviated version of the past screen representation, and its correct output. The output in the example could either be an action originally generated by the LLM during the Derive phase or one that has been modified by the user through the HITL task repair.
In any case, the provided example always demonstrates a correct answer. By providing such examples, we effectively guide the LLM to produce consistent responses based on its memory. Moreover, if the example action is the one that has been corrected by the user, it prevents LLM from repeating the same mistakes.



\section{Implementation}
\red{
Our implementation of \sys{} consists of an Android mobile app and a Python server, which communicates via Wi-Fi. The \sys{} app captures screen representation and injects actions to the smartphone. The Python server manages app memory and processes the Explore, Select, Derive, and Recall phases.
}

\textbf{LLM Agents. }
\sys{} employs multiple LLM agents to cycle through the Explore, Select, Derive, and Recall phases. Each agent is equipped with a model tailored to its operational requirements. For Explore, Select, and Derive agents, which involve mobile screen understanding and multi-step reasoning, we used the most capable GPT-4-turbo language model. For translating user instructions to the high-level task representation and slot-filling subtask parameters, we opted for the faster and more economical GPT-3.5-turbo model. 

\textbf{App Launch. }
When an instruction is given, \sys{} can recommend the most appropriate apps for the given task. It does so by crawling Google Play app descriptions for each app installed on the user's device and storing them in the vector database~\cite{pinecone} using the text-embedding model~\cite{embedding}. Then, when a user gives an instruction, \sys{} retrieves and presents the top three most relevant apps based on their descriptions. When the user makes a selection, \sys{} launches the app and loads its corresponding app memory to proceed with the task.

\textbf{Scrolling. }
When reproducing actions without definite target UI elements (i.e., scroll, sliding), \sys{} checks for its subsequent action's target UIs. If the target is found, it skips all the scrolls in between and directly performs the subsequent action. Otherwise, the system reproduces the scroll actions until it can find the next action's target UI.

\section{Evaluation}
Our evaluation is divided into two sections: comparative study and ablation study. Throughout the evaluation, we used a Google Pixel 6 smartphone.

\subsection{\sys{} Dataset}
\label{sec:dataset}
The main contribution of \sys{} lies in improving performance for repeated tasks and sub-tasks. However, existing datasets (e.g., AITW~\cite{android_in_wild}, DroidTask~\cite{autodroid}, PixelHelp~\cite{mapping}) do not capture this aspect, as they focus on task diversity with a highly independent set of user instructions. 

Therefore, we created a dataset\footnote[1]{The dataset is available at \url{https://github.com/mobile-gpt/MobileGPT}} designed to assess how effectively a system can learn from one instruction and adapt to another. The dataset includes 80 tasks across eight widely used off-the-shelf mobile apps: Google Dialer, Telegram, Twitter, TripAdvisor, Gmail, Microsoft To-Do, Uber Eats, and YouTube Music. Each task is accompanied by two user instructions with different task parameters---totaling \textit{160 user instructions}. For example, the task of "Post Reply" for Twitter comes with two instructions: \textit{"Post reply to Elon Musk's new tweet"} and \textit{"Post reply to Bill Gate's new tweet saying `Reply from \sys{}'."} The dataset's complexity is comparable to the existing dataset, with the following average and maximum step length statistics---\textbf{\sys{}: 5.3(avg)/13(max)}; DroidTask: 4.5/17; AITW: 5.5/14.

\subsection{Comparative Study}
\label{sec:comparitive}
\subsubsection{Experimental Setup.}\hfill\\
\textbf{Baselines.}
We evaluate \sys{}'s performance against two baseline systems: AppAgent~\cite{appagent} and AutoDroid~\cite{autodroid}. AppAgent is a vision-language model (GPT-4-turbo) powered task automator that uses screenshots to execute tasks. AutoDroid is an LLM (GPT-4-turbo) powered task automator that uses text screen representation, similar to \sys{}. The key distinction between \sys{} and the two baseline systems is that \sys{} leverages app memory to learn and recall past executions, thereby optimizing costs and latency.
For \sys{}, we used the GPT-4-turbo model for the Explore, Select, and Derive phases and the GPT-3.5 Turbo model for slot-filling subtask parameters. For each baseline, we used their open-source version available on GitHub~\cite{autodroid_github, appagent_github}.

\noindent\textbf{Dataset.}
We evaluate the performance of three systems using datasets from all three works. After excluding overlapping or deprecated apps and tasks, the total number of apps is 18, and the total number of tasks is 185. This breaks down as follows: AutoDroid dataset (7 apps, 91 tasks), AppAgent dataset (3 apps, 14 tasks), and \sys{} dataset (8 apps, 80 tasks). The average task complexity, measured by the number of steps, is 4.0, 6.6, and 5.3, respectively, and screen diversity, measured by the number of unique app pages, totals 237 (AutoDroid: 82, AppAgent: 39, MobileGPT: 116).

\noindent\textbf{Procedure.}
The evaluation of \sys{} is divided into two stages: \textit{cold-start} and \textit{warm-start}. In the \textit{cold-start}, \sys{} executes the task for the first time and constructs its memory. Then, in the \textit{warm-start}, it utilizes the memory to execute the same task again. Note that AutoDroid and AppAgent run each task only once, as they do not have distinctions between cold-start and warm-start.

We manually monitored each step to identify errors during task execution. When an error occurred, we allowed the system three additional attempts for self-correction. A task was considered successful if the system navigated to the final screen within these three attempts. To ensure a fair evaluation in the comparison study, we did not utilize \sys{}'s HITL task repair feature.

\begin{table}[]
\small
\begin{tabular}{ccccc}
\hline
\multirow{2}{*}{Dataset} & \multirow{2}{*}{\textbf{AutoDroid}} & \multirow{2}{*}{\textbf{AppAgent}} & \multicolumn{2}{c}{\textbf{MobileGPT}} \\ \cline{4-5} 
          &        &      & Cold            & Warm            \\ \hline
AutoDroid & 83.5\% & 80.2\% & \textbf{86.8\%} & 85.7\%          \\
AppAgent  & 27.3\% & 21.4\% & \textbf{50.0\%} & \textbf{50.0\%} \\
MobileGPT & 68.8\% & 60.5\% & \textbf{82.5\%} & \textbf{82.5\%} \\ \hline
Overall   & 74.7\% & 67.4\% & \textbf{82.7\%} & 82.16\%         \\ \hline
\end{tabular}
  \caption{Task success rate across different datasets.}
  \label{tab:compare_accuracy}
\end{table}

\subsubsection{Accuracy.}\hfill\\
\autoref{tab:compare_accuracy} compares the task success rate of each system across different datasets. The success rate of AppAgent dataset is significantly lower because the dataset includes more complex apps (YouTube, Spotify, Google Maps). 

Notably, \sys{}'s \textit{cold-start} demonstrates the highest accuracy across all three datasets, which can be attributed to several factors. First, AppAgent exhibits the lowest task accuracy because it relies solely on screenshot images, which proves inadequate for text-heavy applications such as Gmail, Twitter, and Telegram. Second, AutoDroid represents the app screen in a list of interactable UI elements. However, this overly abbreviated representation often omits crucial information about the screen, such as the hierarchical relationship between UIs and layout UI elements that describe nearby UIs, potentially misleading the LLM to make incorrect decisions.

In contrast, \sys{} retains hierarchical information and minimally omits UI elements when converting the screen to a text representation, preserving as much information as possible from the original Android layout file. Additionally, \sys{}'s method of splitting tasks into subtasks, which resembles the chain-of-thought prompting techniques~\cite{cot, voyager}, contributes to its overall \textit{cold-start} task accuracy. 

\sys{}'s \textit{warm-start} had one additional failure than its \textit{cold-start} due to an error while recalling a task from memory.
Yet, the fact that the \textit{warm-start} accuracy is nearly identical to the \textit{cold-start} demonstrates that \sys{} can recall tasks very successfully using the memory built through cold-start (More details in \autoref{sec:ablation}).

\subsubsection{Cost \& Latency.}
\begin{figure}[t]
    \centering
    \includegraphics[width=0.49\textwidth]{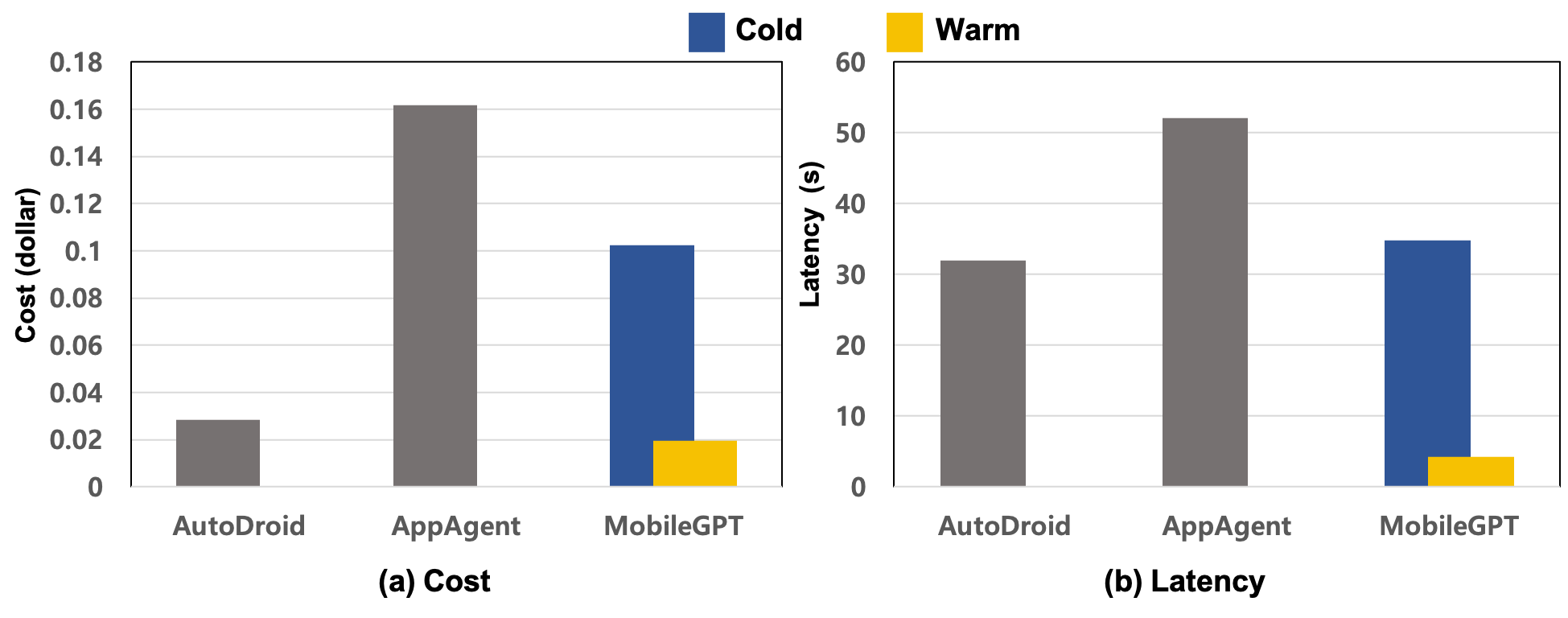}
    \caption{Average task completion cost and latency of AutoDroid, AppAgent and \sys{}}
    \label{fig:compare_efficiency}
\end{figure}

\hfill\\
\autoref{fig:compare_efficiency} illustrates the average task latency and cost\footnote[2]{At the time of evaluation, the LLM costs (per 1K tokens) were: GPT-4-Turbo \$0.01, GPT-3.5 Turbo \$0.003}. Comparisons were made only for tasks successfully completed by all three systems (77 tasks). The number of steps taken for each task was comparable across all systems.

AutoDroid achieves significantly lower costs through its concise screen representation and instruction prompts (i.e., system prompts).
However, note that this cost-effectiveness may come at the expense of accuracy, as overly abbreviated screen representation could omit important details, and LLMs typically perform better with more comprehensive instructions.
Conversely, AppAgent exhibits the highest cost and latency despite the lowest accuracy, demonstrating that relying solely on screenshots is highly inefficient both in terms of accuracy and cost.

\sys{}'s \textit{cold-start} cost and latency fall between AutoDroid and AppAgent. \sys{}'s \textit{warm-start}, however, exhibits the lowest cost and latency, with 87\% and 90\% reduction in latency compared to those of AutoDroid and AppAgent. This high efficiency is largely due to \sys{} having learned the necessary sub-tasks and actions during the initial cold-start stage, leaving only simple tasks such as subtask slot-filling for the \textit{warm-start} stage. We further analyze these benefits in \autoref{sec:ablation}.

Overall, the comparison demonstrates that \sys{} has slightly higher latency and cost during cold starts but becomes significantly faster and cheaper during warm starts. Given the recurrent nature of mobile tasks~\cite{redundant1,redundant2,redundant3,redundant4}, our approach is highly practical as it significantly reduces the burden of common warm-starts, with only a minor increase in that of cold-starts.



\subsection{Ablation Study}
\label{sec:ablation}
In evaluating LLM-based systems, it is crucial to recognize that system performance heavily depends on the prompts used---a system could perform better simply from better prompts. Therefore, isolating the effect of prompts is essential to accurately measure the architectural benefits of such systems.

\subsubsection{Experimental Setup.}\hfill\\
\textbf{Baseline.}
To further analyze \sys{}'s structural design, including its three-phase inference structure (Explore-Select-Derive) and memory, we compared \sys{} against a custom baseline system. This baseline is designed to share the same prompts as \sys{} while following the traditional Derive-only structure employed in prior approaches~\cite{autodroid, droidbot, enabling}. This comparison allows us to measure performance gains attributable solely to \sys{}'s unique structural design.

\noindent \textbf{Dataset. }
In the ablation study, we used the \sys{} dataset, as datasets from other works cannot fully assess \sys{}'s unique ability to learn and recall tasks (see \autoref{sec:dataset}).

\noindent \textbf{Procedure.}
The evaluation is divided into \textit{cold-start} and \textit{warm-start} stages. The \textit{cold-start} executes the first instruction set from each high-level task, and the \textit{warm-start} executes the second instruction set. This procedure assesses how well \sys{} learns from one instruction and adapts to another with different task parameters.

When the system deviates from the correct path, we allow three extra attempts for self-correction. If it fails to correct itself, we mark the task as a failure and manually repair the error using the HITL task repair. After the repair, we let the system continue the task. In the case of \sys{}, The final path after the repair gets stored in the memory and utilized during the \textit{warm-start}.

\noindent \textbf{Offline exploration.}
For \sys{}, we used a random explorer to \textit{explore} each target app before beginning the tasks. we discovered 50 unique app pages for each app, which typically took between 10 to 15 minutes. These app pages accounted for 89.65\% (104 out of 116) of the total app pages required during the evaluation. The remaining app pages, along with those unsuitable for random exploration (e.g., payment page), can be explored offline by monitoring user interactions with the app. The total cost for this random exploration was \$10.78, which is deemed reasonable considering that this preparation is a one-time process.

\begin{figure}[t]
    \centering
    \includegraphics[width=0.49\textwidth]{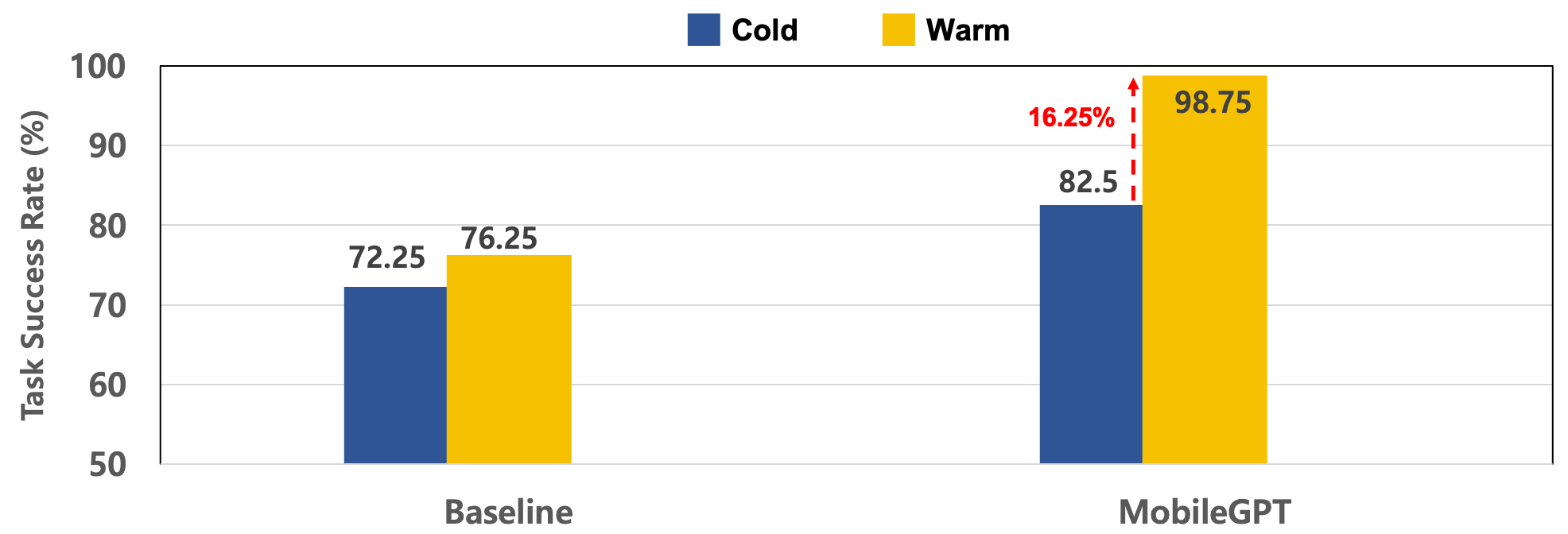}
    \caption{Task success rate of \sys{} and the baseline.}
    \label{fig:ablation_Accuracy}
\end{figure}



\subsubsection{Task Success Rate}\hfill\\
\autoref{fig:ablation_Accuracy} illustrates the task success rate of \sys{} compared to the baseline. Impressively, \sys{} achieves near-perfect accuracy (98.75\%) during the warm-start phase, with a single exception observed in Uber Eats. This improvement over the cold-start accuracy is due to the user’s ability to correct task executions before saving them to memory, ensuring reliability and correctness of the task recall even if LLMs made mistakes during the cold-start. For instance, in Gmail’s \textit{"Recover Deleted Email"} task, \sys{} initially failed to open the More Options Menu because there were two seemingly identical More Options buttons on the screen. However, after "teaching" it which button to click through HITL task repair, \sys{} successfully clicked the correct button during the warm start. Furthermore, \sys{}'s combination of Attribute-based and In-context action adaptation enables precise and consistent recall of learned tasks, even when task parameters and resulting screen content change.

\sys{} consistently outperforms the baseline, even during the cold-start stages. This advantage arises from two factors: \sys{}'s capability to decompose complex tasks into multiple subtasks and its ability to share learned subtasks across different tasks. Notably, when \sys{} makes an error during a subtask and receives a correction, either through self-feedback generation or HTIL task repair, it avoids repeating the same mistake in other tasks involving overlapping subtasks. For example, finding a contact in Telegram requires using the search button instead of scrolling through the contact list. In its initial task, \sys{} attempts to scroll to locate the contact. However, after correcting this approach through self-feedback generation and learning to use the search button, \sys{} subsequently goes directly to the search button for similar tasks, bypassing the scrolling.

In contrast, the baseline system frequently resorts to scrolling through contacts, consistently repeating the same errors across tasks. Furthermore, the baseline sometimes fails or performs inefficiently during the warm start phase, even if it has successfully completed the task during the cold start. This inconsistency highlights the non-deterministic nature of LLMs, which can generate varying outputs based on factors such as task parameters, screen representation, and even the timing of the query.


\begin{table}[]
\small
\begin{tabular}{c|ccc|cl}
\hline
 &
  \multicolumn{3}{c|}{\cellcolor[HTML]{EFEFEF}\textbf{\begin{tabular}[c]{@{}c@{}}Task Learning\\ (cold-start)\end{tabular}}} &
  \multicolumn{2}{c}{\cellcolor[HTML]{EFEFEF}\textbf{\begin{tabular}[c]{@{}c@{}}Task Recall\\ (warm-start)\end{tabular}}} \\ \cline{2-6} 
\multirow{-2}{*}{\begin{tabular}[c]{@{}c@{}} Accuracy\\ (\%)\end{tabular}} &
  \multicolumn{1}{c|}{Explore} &
  \multicolumn{1}{c|}{Select} &
  Derive &
  \multicolumn{1}{c|}{\begin{tabular}[c]{@{}c@{}}Slot\\ Filling\end{tabular}} &
  \begin{tabular}[c]{@{}l@{}}In-context\\ Adaptation\end{tabular} \\ \hline
Phase & \multicolumn{1}{c|}{96.4\%} & \multicolumn{1}{c|}{96.2\%} & 99.1\% & \multicolumn{1}{c|}{99.5\%} & 100\% \\ \hline
Step  & \multicolumn{3}{c|}{95.5\%}                                        & \multicolumn{2}{c}{99.8\%}          \\ \hline
\end{tabular}
\caption{\sys{}'s step and phase accuracy.}
\label{tab:query_accuracy}
\end{table}

\subsubsection{Step Accuracy.}\hfill\\
To further analyze accuracy, we examine LLM queries at each step of the tasks. In \sys{}, each step is broken down into multiple phases: during the cold-start, \sys{} goes through Explore, Select, and Derive phases to learn new tasks. During the warm-start, it employs sub-task slot-filling and in-context action adaptation for recalling learned tasks. A failure in any single phase results in the entire step being marked as a failure. The step accuracy of the baseline is 92.4\%

\autoref{tab:query_accuracy} presents \sys{}'s accuracy across its different phases. \sys{} exhibits a higher step accuracy than the baseline, even during cold-starts. This improvement can be traced to \sys{}'s ability to distribute the reasoning load for each step across multiple phases, similar to how the Chain-of-thought prompting~\cite{cot} improves LLM accuracy by decomposing problems into a series of intermediate steps before reaching the final solution.



When recalling previously learned tasks (\textit{warm-start}), \sys{} demonstrates near-perfect accuracy, with only one slot-filling query failing. Remarkably, \sys{} achieves 100\% accuracy for in-context adaptation. In our evaluation, out of 327 primitive actions, 53 actions (16.2\%) were not adaptable through attribute-based adaptation and therefore required in-context adaptation via LLM. As a result, LLM was able to adapt all 53 actions to new instructions and screens. 


\begin{figure*}[t]
    \centering
    \includegraphics[width=1\textwidth]{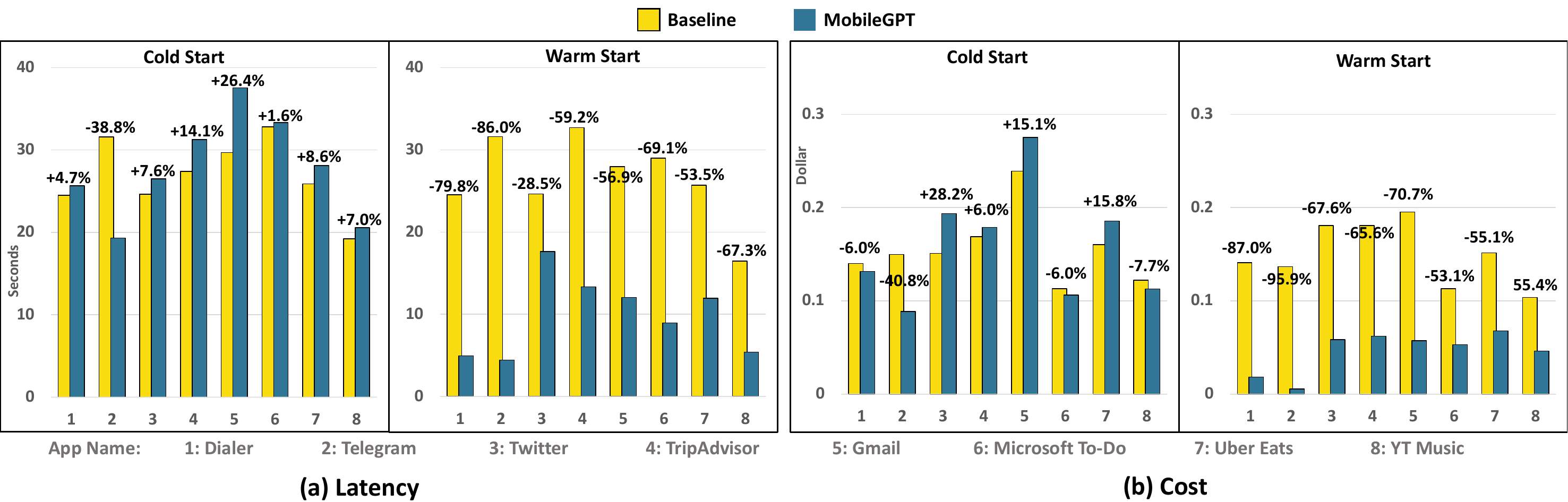}
    \caption{Average task latency and cost of \sys{} and the baseline}
    \label{fig:performance}
\end{figure*}

\subsubsection{Cold-start latency \& cost.}\hfill\\
\autoref{fig:performance} compares the latency and cost between \sys{} and the baseline. During the initial cold start, even with the additional Select phase that \sys{} employs to break down tasks into subtasks, the increase in latency and cost is minimal, averaging only 3.9\% and 0.6\%, respectively.

Moreover, \sys{} even outperforms the baseline in \texttt{Telegram} due to its ability to reuse learned subtasks to be reusable across different tasks. \autoref{tab:hit_rate} displays the average memory hit rate for each app, indicating the percentage of actions retrieved directly from memory without engaging the LLM. Since LLM is the primary source of latency, a higher memory hit rate results in faster task execution. \texttt{Telegram} serves as an excellent example; since many of its tasks involve the common subtask of searching for a specific contact, \sys{} significantly outperforms the baseline, even during the cold-starts. Conversely, \texttt{Gmail} shows a low memory hit rate (6.7\%), as its tasks have few overlapping sub-tasks. While current results suggest that \texttt{Gmail} does not fully leverage our design when learning new tasks (cold-start), we anticipate that both latency and cost will decrease exponentially as more subtasks are accumulated over time.

\begin{table}[]
\small
\begin{tabular}{c|cc}
\hline
\multirow{2}{*}{App Name} & \multicolumn{2}{c}{\textbf{Average Memory Hit Rate}}               \\ \cline{2-3} 
                          & \multicolumn{1}{c|}{\textbf{Cold-Start}} & \textbf{Warm-Start} \\ \hline
Google Dialer             & \multicolumn{1}{c|}{35.9\%}                & 95.8\%                \\
Telegram                  & \multicolumn{1}{c|}{50.1\%}                & 98.6\%                \\
Twitter                   & \multicolumn{1}{c|}{31.7\%}                & 80.1\%                \\
TripAdvisor               & \multicolumn{1}{c|}{29.1\%}                & 89.3\%                \\
Gmail                     & \multicolumn{1}{c|}{6.7\%}                 & 70.6\%                \\ 
Microsoft To-Do           & \multicolumn{1}{c|}{27.8\%}                & 92.6\%                \\ 
Uber Eats                 & \multicolumn{1}{c|}{26.1\%}                & 78.0\%                \\ 
YouTube Music             & \multicolumn{1}{c|}{18.0\%}                & 87.2\%                \\ 
\hline
\end{tabular}
  \caption{\sys{} Average Memory Hit Rate per App.}
  \label{tab:hit_rate}
\end{table}
\normalsize

\subsubsection{Warm start latency \& cost.}\hfill\\
In the warm start stage, where the same high-level task is repeated but with different parameters (i.e., different user instructions), \sys{} dramatically reduces latency and cost, achieving improvements of 62.5\% and 68.8\% over the baseline, respectively. The effectiveness of \sys{}'s memory varies depending on the memory hit rate. The memory hit rate for \textit{warm start} is not always 100\% because some actions still require LLM for its adaptation (i.e., In-context adaptation). Nevertheless, even in the worst case where no actions are adaptable (e.g., Twitter), our evaluation indicates that \sys{} consistently outperforms the baseline.

\subsection{Resource Consumption}
We measured the resource consumption of the \sys{} on mobile devices using the Android Studio profiler tool~\cite{profiler}. The analysis indicated no significant overhead, as the majority of computations are offloaded to the server. The average power consumption of the \sys{} app during task execution was 330.38 mW, comprising CPU at 237.05 mW, memory at 67.15 mW, and WLAN at 26.18 mW, representing 11.78\% of the total power consumption. he average memory consumption was 213 MB, which is only 0.04\% of a typical modern smartphone’s memory capacity (6GB).

\subsection{User Study} 
\textbf{Participants and Study Procedure. }
We recruited 23 participants (16 male, 7 female, mean age 23.2, SD=3.5, range=19-32) through an online community posting, with each participant receiving a compensation of 12 USD. The study consisted of two sessions: the first session evaluated the overall usability of \sys{} in comparison to Samsung Bixby Macro (Programming by Demonstration) and the custom baseline. The second session assessed the usability of \sys{}'s human-in-the-loop task repair mechanisms. Each session lasted 30 minutes. All recruitment and experiments complied with our institution's IRB policies.

In the first session, participants were asked to perform the following three tasks using the three task automation tools: \textit{1) "Find me available hotels in Las Vegas from November 10 to November 15," 2) "Find me hotels in New York," and 3) "Find me restaurants in Las Vegas."} These tasks were designed to assess the usability of the automation tools across three scenarios: \textit{i)} executing a completely new task, \textit{ii)} repeating the same task with different parameters, and \textit{iii)} executing a new but similar task. After each task, participants rated the usability of each tool on a 7-point Likert scale. Upon completing all tasks, participants ranked the tools based on their preferences and indicated their willingness to use each tool in real-world scenarios.

In the second session, participants were introduced to \sys{}'s repair mechanisms, which allow users to correct mistakes made by the LLM. After a brief tutorial, participants were tasked with executing the instruction \textit{"Modify Tom's phone number to 123-456-789"} using \sys{}. During the study, \sys{} was configured to make predefined errors covering all three areas of repair: Explore, Select, and Derive. Participants were not informed in advance about the specific mistakes the system would make. Following the task, they rated the usability of the repair mechanism using the System Usability Scale (SUS)~\cite{sus}. Additionally, they evaluated the necessity and effectiveness of the repair mechanism, along with their accuracy tolerance for a task automator with and without the repair mechanism.

\begin{table}[]
\small
\begin{tabular}{c|ccc}
\hline
             & \multicolumn{3}{c}{Overall Usability (7-point-scale)}         \\ \hline
Scenario     & \textbf{PbD} & \textbf{Baseline} & \textbf{\sys{}} \\ \hline
New Task     & 3.2          & 4.4               & 4.7                \\
Repeat Task   & 3.0          & 4.3               & 6.1                \\
Similar Task & 3.0          & 4.3               & 5.7                \\ \hline
\end{tabular}
\caption{Usability score for different task scenarios.}
\label{tab:overall_usability}
\end{table}

\textbf{Session 1: Overall usability. }
\autoref{tab:overall_usability} shows the usability scores for each task automation tool across scenarios. Throughout the scenarios, participants generally found PbD inefficient because they needed to re-create the macro script even at the slightest change in the screen or instructions, and the baseline as convenient but too slow. In contrast, \sys{} initially had a usability score similar to the baseline in the first scenario, but showed significant improvement in the subsequent tasks, with Mann-Whitney U Test results of (U=216.5, p=0.26), (U=24.0, p=4.8$e^{-8}$), and (U=67.5, p=6.5$e^{-6}$), respectively---\textit{"(\sys{}) performs the task accurately at a fairly high speed. Possibly faster than doing it by hand. Seems to have good generalizability" (P4)}, \textit{"It was a little sluggish for things that I hadn't done before, but it was still faster than the baseline and I figured it would get faster with more training" (P7)}. This trend suggests that \sys's ability to quickly reproduce tasks and adapt learned sub-tasks to related tasks greatly enhances its usability. This is corroborated by the post-survey results, where all but one participant favored \sys the most, and all participants expressed willingness to use \sys in real-world applications, compared to 35\% and 30\% for the baseline and PbD tools, respectively.


\textbf{Session 2: Human-in-the-loop Task Repair. }
\sys{}'s task repair mechanism received an average score of 64.6 (std=16.9) on the System Usability Scale, indicating it as \textit{"ok"} user-friendliness~\cite{sus_score}. While there is room for improvement, the score is considered acceptable given the high complexity of the repair mission (repairing all three phases within a single task) and the participants' general unfamiliarity with the concept of autonomous agents making errors.

In the post-survey, participants rated the necessity and effectiveness of the repair mechanism highly, with scores of 4.7 and 4.8 out of 5, respectively. Participants also indicated that the acceptable accuracy threshold for a task automator without a repair mechanism is 96\% on average, whereas, with a repair mechanism, this threshold drops to 84\%. This indicates that participants are more lenient with accuracy expectations when a repair mechanism is available, underscoring its value in task automation.

\section{Related Work}
\textbf{Using LLM in UI Task automation. }
Recently, there have been several efforts to leverage LLMs for task automation~\cite{autodroid, droidbot, enabling, appagent, webgpt}, capitalizing on their ability to comprehend and perform tasks without requiring prior training or user demonstrations. Noteworthy examples include AutoDroid~\cite{autodroid} and AppAgent~\cite{appagent}, which enhance LLMs with app-specific knowledge to improve their understanding of mobile applications and boost task performance accuracy. Despite these advancements, the practical deployment of these approaches in real-world task automation remains uncertain due to the inherent unreliability and high costs associated with LLMs. \sys{} tackles these challenges by integrating human-like app memory with LLMs, allowing for the effective reuse of learned subtasks, which significantly reduces the number of LLM queries, thereby optimizing latency and cost.



\textbf{Macro Mining.} 
Another line of work closely related to \sys{} is macro mining~\cite{glider, andenv, data-driven, mapping, macro_mining}, which aims to generate macro scripts without human intervention. For instance, Li et al.~\cite{macro_mining} utilize LLM to discover sub-tasks within app traces and combine them to create task-oriented macros. However, a common challenge in using macros is the generalization of actions. Existing approaches tackle this challenge through embedding-based screen matching and attribute-based UI matching. However, these methods often struggle when screen similarity is ambiguous or when UIs lack critical attributes. In contrast, \sys{} effectively addresses these issues by using sub-task-based screen classification and in-context action adaptation, enabling more robust and reliable task automation.

\textbf{Caching LLM responses. }
Fundamentally, \sys{} is built upon the principles of caching LLMs' responses. Similar research efforts include caching frequently occurring token states for faster inference~\cite{prompt_cache}, caching LLM responses to handle similar queries~\cite{cache_embedding}, and training smaller models to replicate cached responses~\cite{cache_distill, cache_me}. However, these existing studies primarily focus on the use of LLMs in chatbot applications. To the best of our knowledge, \sys{} is the first to explore caching within the context of an LLM-based task automator.

\textbf{LLM-based Autonomous Agent in other fields. }
LLMs demonstrate significant potential in automating human tasks. Specifically, it can derive the most appropriate actions for specific tasks when provided with a proper action space (e.g., tools, function calls, APIs). Consequently, researchers across various fields have attempted to expand the capabilities of LLMs by integrating UI interactions~\cite{autodroid, droidbot, android_in_wild, webgpt}, programming tools~\cite{tptu, hugginggpt}, APIs~\cite{gorilla, apibank, restgpt, taskmatrix, toolfermer, toolllm}, and games~\cite{voyager, minecraft, generative}. \sys{} is unique in its ability to generate a higher-level action space (i.e., sub-tasks) on its own and utilize them for different goals (i.e., user instruction). This approach can be extended to other fields, as many digital tasks share common and recurring sub-tasks.

\section{Discussion}
\textbf{Security \& Privacy. }
Screen representations can contain personal information, such as names and phone numbers, posing privacy risks when transmitted to the LLM. To mitigate this, we can employ a Personal Identifier Information (PII) Scanner~\cite{pii1, pii2} to detect personal data within prompts and replace it with non-private placeholders. Additionally, certain actions---e.g., agreeing to terms of service or confirming a payment—should be performed with user supervision. To address this, we provide the LLM with the option to select a \textit{`get user confirm'} action during the Derive phase, ensuring that any potentially risky actions are identified and user confirmation is requested before proceeding.

\textbf{Sharing App Memory. }
\sys{} currently stores memory on a local basis, meaning each device has its own version of app memory. To enhance this, the memory can be shared or crowd-sourced to create a large-scale app memory. This approach can effectively eliminate the need for each user to learn tasks individually, making the cold-start learning stage even less common. To accommodate the varying devices, app memory can be categorized based on the device form factor, as the app interface varies depending on the form factor (e.g., smartphone or tablet device). The diversity of the device resolution is not a concern, as \sys{} identifies UI elements using their hierarchical structure (i.e., index), not their pixel coordinates.  

\textbf{Cross App Task execution. }
Users may give instructions that involve multiple apps. For example, the instruction "Check my bank account, and if it has over 10 dollars, order me a pizza" requires interaction with both a banking app and a food delivery app. To accommodate such scenarios, \sys{} can be extended to maintain a global dataset of known tasks across apps. \sys{} can then use this to plan out which task to execute in which order to fulfill the user instruction.

\textbf{Unsupported Apps. }
The current implementation of \sys{} does not support mobile apps that lack text representation of their screen layouts, such as those using third-party UI engines (e.g., Flutter, Web Apps) or screens dominated by images (e.g., maps, camera views). To address this limitation, we can leverage screen-to-text translation models~\cite{screen_recognition, webui, pix2struct, spotlight} or vision-language models (VLMs)~\cite{llava, visionllm, gpt4vision} to process both the screenshot and the text-based screen representation, enabling \sys{} to function effectively across a wider range of applications.

\textbf{Using Vision Language Model. }
We have performed a feasibility study on using a VLM (GPT-4-turbo) within \sys{}. We have found that \sys{} achieves higher accuracy when given both the screenshot and the screen representation. One limitation, however, is the increased latency of the VLM compared to text-only LLMs. Nonetheless, because \sys{} can minimize the number of queries through memory caching, the higher per-query latency leads to a more pronounced benefit of \sys{}'s design.
\section{Conclusion}
We introduced \sys{}, a novel LLM-powered mobile task automator that enhances the efficiency and reliability of task automation by emulating human cognitive processes. We anticipate that \sys{} will strengthen the integration of intelligent automation into everyday technology use.

\begin{acks}
This work was supported by IITP (RS-2023-00232728), TIPA (TIPS 00262147, RS-2024-00447529), K-Startup (20144069), and the National Research Foundation of Korea(NRF) grant (RS-2024-00347516, RS-2023-00222910).
\end{acks}

\bibliographystyle{ACM-Reference-Format}
\bibliography{references}
\end{document}